\documentstyle[sprocl,epsfig]{article}

\def\Journal#1#2#3#4{{#1} {\bf #2}, #3 (#4)}

\def\NPB{{\em Nucl. Phys.} B}
\def\PLB{{\em Phys. Lett.}  B}
\def\PRL{\em Phys. Rev. Lett.}

\def\EPJ{{\em Eur. Phys. J.} C}

\def\be{\begin{equation}}
\def\ee{\end{equation}}
\def\bea{\begin{eqnarray}}
\def\eea{\end{eqnarray}}
\def\alphas{\alpha_{s}}
\def\gev2{GeV$^2$}

\begin{document}

\title{TESTS OF QCD: SUMMARY OF DIS 2000}

\author{ROBERT BLAIR}
\address{Argonne National Laboratory, Argonne, Illinois 66439, USA}

\author{G\"UNTER GRINDHAMMER}
\address{Max-Planck-Institut f\"ur Physik (Werner-Heisenberg-Institut),\\
F\"ohringer Ring 6, D-80805 M\"unchen, Germany}

\author{MICHAEL KLASEN}
\address{II.\ Institut f\"ur Theoretische Physik, Universit\"at Hamburg,\\
Luruper Chaussee 149, D-22761 Hamburg, Germany}

\author{MICHAEL KR\"AMER}
\address{Department of Physics and Astronomy, University of 
 Edinburgh,\\ Edinburgh EH9 3JZ, Scotland}

\maketitle\abstracts{This summary of the working group 2 of DIS 2000
encompasses experimental and theoretical results of jet physics, open
and bound state heavy flavour production, prompt photon production,
next-to-leading order QCD calculations and beyond, instantons,
fragmentation, event shapes, and power corrections, primarily from
deep-inelastic scattering and photoproduction at \mbox{HERA}, but also
from the \mbox{LEP} and \mbox{Tevatron} colliders.}

\section{Jets in deep-inelastic scattering}

Sch\"orner, P\"oschl, Chapin, and Caron reported on measurements of
inclusive jet (1+1)~\cite{schoerner}, dijet
(2+1)~\cite{poeschl,chapin}, and three jet (3+1)~\cite{caron}
production by H1 and ZEUS, scanning an enormous region of the
kinematic plane and of the jet phase space at \mbox{HERA}.  They
provide therefore an important and demanding test of the adequacy of
the current QCD calculations in next-to-leading order (NLO) of the
strong coupling parameter in these regions.

In the photon virtuality $Q^2$ the measurements cover the range from
5~\gev2 to 10$^4$~\gev2 , in Bjorken-$x$ from 10$^{-4}$ to 0.3, and in
$\xi = x (1+M^2_{jj}/Q^2)$ from 5$\times 10^{-3}$ to 0.1, where $\xi$
in leading order (LO) of the strong coupling parameter is the momentum
fraction of the gluon or quark emitted by the proton.  Jets are
defined using the inclusive $k_{T}$-algorithm in the Breit
frame.~\cite{poeschl} In this frame the transverse energy of a jet is
dominated by QCD processes.  In the jet phase space the measurements
span the range in the transverse energy of the jets in the Breit frame
from $ 5 < E_{T} < 100$~GeV, in $E^{2}_{T}/Q^2$ from $0.5$ to $100$,
in rapidity from forward to central and backward jets in the
\mbox{HERA} frame, and in the three jet invariant mass $M_{3j}$ up to
$100$~GeV.

The comparison of these jet measurements to QCD calculations in NLO
allows to draw the following conclusions: 1) Calculations using $Q^2$
as renormalization scale can describe almost all of the data at the
prize of large uncertainties at low $Q^2$, when varying the scale by a
factor of 1/4 to 4.  2) When using $E_{T}^2$ as the relevant scale,
good agreement with the data is achieved for large jet $E_{T}$ ($E_{T}
> 20$~GeV)~\cite{schoerner} and for large $Q^2$ ($Q^2 >
50$~\gev2)~\cite{poeschl,chapin}.  For intermediate values of the
ratio $E^{2}_{T}/Q^2$, between 1 and 50, the calculations are below
the data~\cite{schoerner,chapin}.  The forward jet cross section shown
by H1~\cite{schoerner}, requiring $x_{jet} > 0.035$ and $1.5 <
\eta_{lab} < 2.8$, but without demanding that $E_{T}^2 \approx Q^2$ as
done in previous such analyses by H1 and ZEUS~\cite{fwd-jets}, rises
more strongly than the NLO prediction for decreasing $x$ down to
$10^{-4}$.  With $E_{T}^2$ as the renormalization scale the
uncertainties in the predicted cross sections are however much smaller
than for the scale $Q^2$.  3) Typically the discrepancies between data
and NLO calculations are large, where the corrections from LO to NLO,
the hadronisation corrections, and the renormalization scale
uncertainties are large.  4) In a large part of the phase space the
theoretical uncertainty due to the renormalization scale are much
larger than the experimental errors.  The latter are dominated by the
uncertainty of the hadronic energy scale of the calorimeters of both
H1 and ZEUS.

\begin{figure}[htb]
\noindent
\begin{minipage}[htb]{.46\linewidth}
\centering
\epsfig{figure=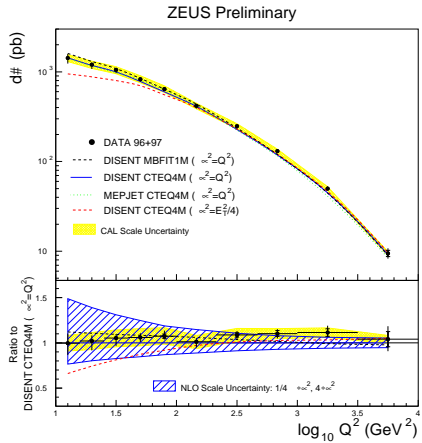,height=58mm,width=\linewidth}
\caption[]{\label{fig:chapin-a}
  Dijet cross section as a function of $Q^2$ compared to NLO QCD
  calculations for the renormalization scales $Q^2$ and $E_T^2/4$
  using DISENT and MEPJET~\cite{Doyle:1999jc}.  For the ratio of data
  to NLO an uncertainty estimate for the scale $Q^2$ is given by the
  shaded band.~\cite{chapin}} 
\end{minipage}\hfill
\begin{minipage}[htb]{.46\linewidth}
\centering\vspace*{-6mm}
\epsfig{figure=chapin-d.eps,height=58mm,width=\linewidth}
\caption[]{\label{fig:chapin-d}
  NLO QCD scale uncertainty estimates for the DIS dijet cross section
  as obtained with the DISENT~\cite{Doyle:1999jc}
  program.~\cite{chapin}}
\end{minipage}  
\end{figure}

The NLO QCD scale uncertainty estimates for the DIS dijet cross
section as a function of $Q^2$ and $<E_T>^2/Q^2$ are exemplified in
Figs.~\ref{fig:chapin-a} and~\ref{fig:chapin-d} respectively.  In
Fig.~\ref{fig:chapin-a} the scale uncertainty can be seen to grow
rapidly for decreasing $Q^2$.  In the lower $Q^2$ region it is much
larger than the experimental error.  In Fig.~\ref{fig:chapin-d}, in
the region where $E_T^2 > Q^2$ the difference between the NLO
predictions using $\mu_r^2 = Q^2$ and $\mu_r^2 = E_T^2/4$ is larger
than the respective estimated scale uncertainties.  Clearly
next-to-next-to leading order (NNLO) and also resummed calculations
are needed to make further progress.

Forward jet cross sections at very small $x$ are supposed to be a
clean test of BFKL or CCFM dynamics.  The latter should be {\em the
way} to describe low-$x$ final states to leading double-log accuracy.
L\"onnblad discussed the implementation of the CCFM equation in
different Monte Carlo generators.~\cite{lonnblad} Two of the programs
implementing CCFM, SMALLX~\cite{Doyle:1999jc} and the new complete
final state Monte Carlo generator CASCADE~\cite{Doyle:1999jc} are able
to describe the forward jet cross sections.  However they do so only,
if the so-called consistency constraint $k^2_{ti} > z_{i}q^2_{ti}$,
which has been found to be necessary in BFKL dynamics to account for
kinematic effects, is not applied.  In addition, the results are found
to be sensitive to the treatment of non-singular terms in the gluon
splitting function.  However, no complete calculation including these
terms has been done, and therefore no firm conclusions can be drawn.

\section{Jets in $\gamma$p, $\gamma^{*}$p, and $\gamma\gamma$}

A photon is not just a photon and not just a hadron.  In LO QCD it has
two components, a direct one, which couples electromagnetically to one
of the partons of the other beam, and a resolved one.  In the latter
it fluctuates into a $q\bar q$ or even more complicated partonic
state.  In jet production the probing scale is given by the $E_{T}^2$
of the jets and the `size' of the photon by $Q^2$.  Therefore, when
$E_{T}^2 \gg Q^2$ the structure of the photon may be resolved.

Maxfield discussed the assumptions and problems encountered in the
measurement of cross sections for low $E_{T}$ jet production by real
and virtual photons and the extraction of the effective photon parton
densities by H1.~\cite{maxfield} Measurements at low $x_{\gamma}$, the
fractional momentum of the parton in the photon, require measuring
low-$E_{T}$ jets.  This demands a good experimental understanding and
simulation of the pedestal transverse energy due to multiple
interactions of the remnant partons of the photon and the proton in
addition to the $E_{T}$ of the jets from the hard subprocess.

\begin{figure}[htb]
\vskip -0.cm
\hskip 1.75cm
\epsfig{figure=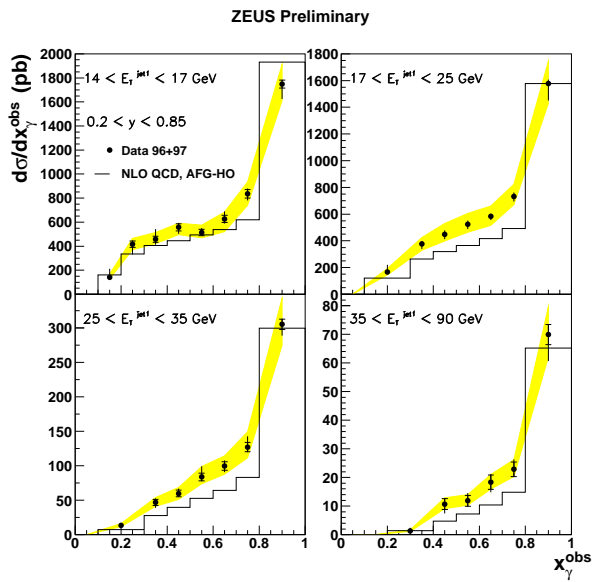,height=8cm}
\vskip -0.2cm
\caption[]{\label{fig:wing}
  Dijet photoproduction cross section measurement in
  $x_{\gamma}^{\mbox{obs}}$ in slices of $E_T^{\mbox{jet1}}$ compared
  to the NLO QCD prediction.  The band around the points displays the
  error due to the uncertainty associated with the calorimeter energy
  scale.~\cite{wing}}
\vskip -0mm
\end{figure}

The photoproduction of dijets as a function of $E_{T}$ has been
measured by H1 at high $E_{T}$ up to about 90 GeV and is in good
agreement with NLO QCD calculations.~\cite{wing} A ZEUS measurement of
the dijet cross section for lower $E_{T}$ as a function of
$x_{\gamma}$, also presented by Wing, shows the data overshooting the
NLO prediction by about 50\% for $x_{\gamma} < 0.8$, see
Fig~\ref{fig:wing}, which is considered to be due to inadequacies of
the current parameterizations of the photon parton densities at scales
of about $E_{T}^2 \approx 200$~\gev2.~\cite{wing} Improved photon
parton densities should be expected from fits including this new
\mbox{HERA} data.

Surrow showed results from OPAL at \mbox{LEP} on dijet production for
the case of quasi-real photons from both colliding beams, i.e. both
photon virtualities are almost zero.~\cite{surrow} They are well
described by NLO QCD calculations.

\section{Measuring $\alphas$ using jets in deep-inelastic scattering}

The extraction of the strong coupling parameter $\alphas$ with
increasing precision is an important task of the \mbox{HERA}
experiments.  A new result from ZEUS, using measurements of the dijet
cross section and of the dijet rate in bins of the photon virtuality
$Q^2$ was presented by Hadig.~\cite{hadig} In their analysis the
parton density functions (pdfs) are taken from global fits of other
measurements.  For the first time, the error on $\alphas$ due to the
uncertainty in the pdfs is estimated including their
correlations~\cite{botje}.  From the dijet cross section they obtain
\begin{eqnarray*}
    \alphas(M_{Z}) & = & 0.1186 \pm 0.0019 \,{\rm (stat.)} \:
    ^{+0.0020}_{-0.0007} \,{\rm (exp.)} \: ^{+0.0035}_{-0.0033} \,{\rm 
    (E-scale)} \\
                   &   & ^{+0.0048}_{-0.0038} \,{\rm (ren.scale)} \, 
    {\bf \pm 0.0031 \,{\rm (pdf)}} \pm 0.0005 \,{\rm (hadr.)}
\end{eqnarray*}
and from the dijet rate
\begin{eqnarray*}
    \alphas(M_{Z}) & = & 0.1166 \pm 0.0019 \,{\rm (stat.)} \:
    ^{+0.0023}_{-0.0005} \,{\rm (exp.)} \: ^{+0.0036}_{-0.0034} \,{\rm 
    (E-scale)} \\
                   &   & ^{+0.0050}_{-0.0042} \,{\rm (ren.scale)} \, 
    {\bf \pm 0.0012 \,{\rm (pdf)}} \pm 0.0005 \,{\rm (hadr.)} \, .
\end{eqnarray*}

It can be seen that the uncertainties due to the pdfs partially cancel
when using the dijet rate.  The extracted values have an experimental
precision which allows to have an impact on the world average of
$\alphas$.  The major errors which need to be reduced in the future
are the theory error due to the renormalization scale followed by the
experimental error due to the hadronic energy scale of the
calorimeter.

\section{Azimuthal asymmetries in deep-inelastic scattering}

A basic prediction of perturbative QCD, which is however difficult to
verify experimentally, is the distribution of the azimuthal angle
$\Phi$ between the lepton scattering plane and the hadron production
plane, defined by the exchanged virtual photon and an outgoing hadron.
For parton production this azimuthal dependence has the form
\begin{equation}
    d\sigma / d\Phi = A + B\cos\Phi + C\cos 2\Phi, \nonumber
\end{equation}
resulting from the polarisation of the exchanged photon.  Brook
reported on such a measurement by ZEUS using single hard
particles.~\cite{brook} The measured value for $<\cos\Phi>$ is found
to be negative, and the value for $<\cos 2\Phi>$, measured for the
first time, is positive and increases with the transverse momentum of
the particles.  Both results agree, within large statistical errors,
with QCD calculations.

\section{Search for instantons}

Instantons are a fundamental theoretical prediction of QCD. They exist
in QCD and weak interactions and are associated with tunneling
transitions between topologically non-degenerate vacua.
Instanton-induced processes in QCD become calculable and perhaps
observable in \mbox{DIS} due to the presence of a hard scale.  Cross
sections and characteristic event signatures can be calculated using
the Monte Carlo generator QCDINS~\cite{Doyle:1999jc} and were
discussed by Schrempp, together with theoretical uncertainties on the
event topology and the rates.~\cite{schrempp}

Mikocki from the H1 collaboration presented first results on a
dedicated search for instanton-induced processes in
\mbox{DIS}.~\cite{mikocki} They are based on some of the expected
characteristics of the hadronic final state of such processes.
Choosing simple cuts on three observables, requiring highest
background reduction ($\approx$ 0.1\%) while keeping an efficiency of
$\approx$ 10\% for instanton-induced processes, 549 events are
observed in the data.  For the background, i.e. standard \mbox{DIS},
$435^{+36}_{-22}$ events are estimated using the leading order matrix
elements and a parton shower model, which is part of
RAPGAP~\cite{Doyle:1999jc}, and $363^{+22}_{-26}$ using the colour
dipole model, as implemented in ARIADNE~\cite{Doyle:1999jc}.  The
errors on the standard \mbox{DIS} background include only the
experimental systematic errors and no model uncertainties.  An excess
in data over standard \mbox{DIS} events is seen which is qualitatively
compatible with the expected instanton signal.  However, the size and
shape of this excess is at the level of the discrepancy between the
standard QCD models themselves.

\section{Status of higher-order calculations}

QCD predictions in leading order (LO) of perturbation theory suffer
generally from large higher-order corrections and scale uncertainties.
They depend on the renormalization scale in the strong coupling
constant $\alpha_s$ and on the factorization scales in the parton
densities and fragmentation functions.  At next-to-leading order
(NLO), the renormalization and factorization scales appear also
explicitly in the hard cross section, thus reducing the scale
dependence considerably and making the theoretical predictions more
reliable.

NLO QCD calculations for HERA physics have been performed for jet
production in \mbox{DIS}, real and virtual photoproduction, heavy
flavour production in \mbox{DIS} and photoproduction, prompt photon
production in \mbox{DIS} and photoproduction, and inclusive hadron
photoproduction (for references see
e.g.~\cite{Doyle:1999jc,Grindhammer:2000qn}).

The improved performance of \mbox{HERA}, the upcoming luminosity
upgrade, and the better understanding of the H1 and ZEUS detectors
contribute to reduced statistical and systematic errors on
experimental measurements, with the exception of the still substantial
energy scale uncertainty.  Consequently, comparisons between
experimental data and NLO QCD predictions are more and more dominated
by theoretical scale uncertainties, limiting the discriminating power
of the data for $\alpha_s$ and parton density determinations.  For
many observables it appears thus mandatory to improve the theoretical
predictions beyond NLO. The first option consists in calculations in
next-to-next-to-leading order (NNLO) of QCD.

Glover reported that the last three years have seen significant
progress in this technically difficult field: Double unresolved tree
amplitudes and single unresolved one-loop amplitudes have been
calculated in the infrared limit, and many massless two-loop integrals
have been evaluated, the most difficult ones being the double-box
diagrams.~\cite{Glover}

The remaining challenges are to integrate out the infrared behavior of
the unresolved cross sections and to analytically evaluate the
two-loop matrix elements.  It might still take several years before
the first NNLO Monte Carlo program becomes available.

\section{New results from resummation}

An alternate route is to resum QCD corrections from gluon radiation in
the soft (small transverse momentum $q_T$) and threshold (large $x_T =
2 q_T / \sqrt{s}$) regions of phase space.  Resummation techniques
rely on the observation that the perturbative series exponentiates.
They have been applied to $e^+e^-$ annihilation, Drell-Yan production,
and very recently also to \mbox{DIS}, where more work is needed.  For
photoproduction resummed calculations are not available.

Data for $q_T$ distributions in semi-inclusive \mbox{DIS} do not
exist.  However, the rapidity distribution of the transverse energy
flow has been measured by H1~\cite{h1-etflow}, and the rapidity $\eta$
can be related to $q_T$ in the Breit frame if the photon virtuality
$Q$ is fixed ($q_T=Q e^\eta$).  Nadolsky showed that the data compare
favourably to the resummed predictions.~\cite{Nadolsky:2000zv}

Resummed predictions for \mbox{DIS} event shape variables like thrust
have been calculated by Dasgupta, but have not yet been confronted
with data.~\cite{Dasgupta:2000bs} When compared to NLO thrust
distributions, they show a constant difference with respect to
DISASTER++~\cite{Doyle:1999jc} as expected, but a thrust-dependent
difference with respect to DISENT~\cite{Doyle:1999jc}.

\section{Prompt photon production}

A physical process where resummation plays a very important role is
prompt photon production.  Prompt photons could help to constrain the
gluon density in the proton and the photon, since initial gluons
generally dominate the production cross sections by 80\% or more.
Unfortunately, this possibility has been put on hold since the
transverse momentum distributions of prompt photons in fixed target
(e.g.\ E706~\cite{Apanasevich:1998hm}) and collider (e.g.\
UA2~\cite{Alitti:1993kw} and CDF~\cite{Blair}) experiments differ
widely from the perturbative NLO expectations.~\footnote{This is not
the case for off-shell photons, which might therefore be better suited
to determine the unpolarized and polarized gluon densities in the
proton.}

While threshold resummation reduces the scale dependence of the
theoretical prediction and improves the description of the E706 data
at large $q_T$, it does not increase the cross section at small $q_T$.

It has been speculated that intrinsic $q_T$ effects could be
responsible for the discrepancy meaning that the partons in the
colliding hadrons are not exactly collinear.  Phenomenologically this
can be motivated by re-expressing the delta-function $\delta(q_T)$ as
a Gaussian exponential function or using parton densities which are
not integrated over $q_T$.  These considerations must be put on firm
theoretical ground: Recoil effects have to be taken into account, and
the modified parton distributions have to be related to the usual ones
which are used in other scattering processes.

\begin{figure}[htb]
\vskip -0.cm
\hskip 2.4cm
\epsfig{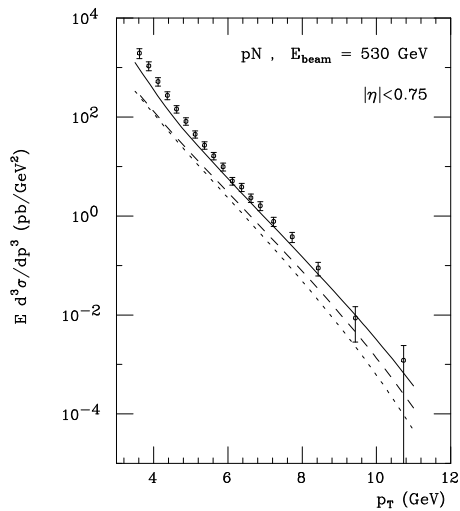}
\vskip -0.2cm
\caption[]{\label{fig:vogelsang}
  $Ed^3 \sigma_{{\rm pN}\to\gamma X}/dp^3$.  The dotted line
  represents the full NLO calculation, while the dashed and solid
  lines respectively incorporate pure threshold resummation and the
  joint threshold and transverse momentum
  resummation.~\cite{Laenen:2000de}}
\vskip -0mm
\end{figure}

A possible solution, reported on by Vogelsang, consists in the form of
a $q_T$-profile function, which also permits to resum the low and high
$q_T$ regions simultaneously.~\cite{Laenen:2000de} Although the
numerical result still lies somewhat below the E706 data at low $q_T$,
the discrepancy is significantly reduced, see
Fig.~\ref{fig:vogelsang}.
A drawback of this (as of any resummed) approach is that it is
necessary to match the resummed and perturbative predictions at a
fixed value of $q_T$ which is {\it a priori} undetermined.

Prompt photon production has also been measured by ZEUS at \mbox{HERA}
in photon-proton scattering and was reported on by
Lee.~\cite{Breitweg:2000su} The measurements cover the range in
transverse momentum $p_{T}^{\gamma}$ from 5 to 15 GeV and in
pseudorapidity $\eta^{\gamma}$ from -0.7 to 0.9.  The $p_{T}^{\gamma}$
distribution agrees very well with NLO QCD predictions, shedding doubt
on the universality of intrinsic $q_T$ effects.  However, the
experimental errors are still dominated by statistics.  An update of
the analysis is in progress.  It is also important to keep in mind
that the colliding photon at \mbox{HERA} has a much smaller transverse
size than a proton (or even a nucleon) in hadron collisions, which
leads to smaller intrinsic $q_T$ effects at least on the photon side.
The NLO calculations describe the $\eta^{\gamma}$ distribution only
well in forward (proton) direction, but are below the data in the
backward region.  This would seem to indicate an inadequacy of the
current photon parton densities.

\section{Non-perturbative hadronization corrections}

Both NNLO and resummed perturbative calculations are furthermore
subject to non-perturbative corrections.  These arise, e.g., when
partons hadronize into jets, and they can become very important in
certain regions of phase space like the backward region of jet
photoproduction.

A possible remedy is the combination of perturbative predictions with
parton showers and hadronization from existing Monte Carlo models.
These models require a well-defined colour flow for the underlying
hard scattering matrix element which poses a problem for interference
terms.  In a first approximation they can be split up and added to the
squared matrix elements.  Furthermore, beyond LO one has to be careful
not to doublecount emission of real partons in the higher order matrix
element and in the parton shower.  The two can be separated by cutting
on the invariant mass of the parton pair.

\section{Fragmentation functions}

The inclusive cross section for the production of single charged
hadrons is expressed in terms of universal fragmentation functions
(FF) which incorporate the long-distance, non-perturbative physics of
the hadronization process.  The scale dependence of the fragmentation
functions is governed by an evolution equation, similar to the
evolution equation for parton densities.  The initial condition for
the evolution equation, however, is not calculable in perturbation
theory and must at present be taken from experiment.

Two new analyses to extract fragmentation functions for charged
hadrons from $e^+e^-$ annihilation data have been presented by
P\"otter and Kretzer.~\cite{poetter,kretzer} Both analyses have been
performed at leading and next-to-leading order in QCD. The fitted
fragmentation functions describe all data sets within their respective
errors.  A comparison between the two independent FF extractions
reveals good agreement when the sum over all flavours is taken.
However, significant differences are found for the individual flavour
fragmentation functions into $\pi^{\pm}$ and $K^{\pm}$, which are not
well constrained by the data.~\cite{kretzer}

Using experimental measurements of inclusive single hadron production
at different energies one can extract the strong coupling $\alpha_s$
from the scaling violations in the fragmentation functions.  A new
$\alpha_s$ determination with a competitive error and a value
consistent with the world average has been presented at this
workshop.~\cite{poetter}

\section{Inclusive particle production and MLLA}

Safonov presented results on jet fragmentation, i.e. inclusive
momentum distribution and multiplicities in jets, from
CDF.~\cite{safonov} He demonstrated that calculations in the
modified-leading approximation (MLLA) and assuming local parton hadron
duality can provide a simple and compact description of the data.
Milstead reported on inclusive particle production in the Breit frame
in \mbox{DIS} as studied by H1 and ZEUS.~\cite{milstead} Here a
consistent description of various moments of the fragmentation
function by MLLA has not yet been achieved.

\section{Event shapes and power corrections}

The first new ZEUS measurements of event shapes were reported by
Wollmer along with those from H1 by Rabbertz.~\cite{rabbertz_wollmer}
These measurements allow for an extraction of $\alpha_s$ and another
parameter $\alpha_0$ that characterizes the behavior of the strong
coupling in both the high energy and low energy (non-perturbative)
regimes.  Several event shapes are considered, but there is a rather
wide spread in results.  Biebel reviewed the measurements from
\mbox{LEP}~\cite{biebel} and Dasgupta the theoretical
situation~\cite{Dasgupta:2000bs} of power corrections.  The conclusion
so far is that there is still a need for deeper understanding of the
experimental and theoretical uncertainties associated with these
measurements.

\section{Photoproduction of charm and beauty}

The production of heavy flavours provides new opportunities to study
the dynamics of perturbative QCD and to extract information on the
proton and photon structure. \mbox{HERA} has provided a wide spectrum
of $D^*$ measurements and first results on $D_s$ mesons by ZEUS were
reported by Gladilin.~\cite{gladilin} It is not guaranteed {\it a
priori} that charm cross sections can be reliably predicted in
perturbation theory.  The NLO calculations are indeed plagued by large
uncertainties and they tend to underestimate the $D^*$ cross section,
in particular in the forward direction and at low $x_\gamma$ where
resolved photon processes contribute.  Potentially large
next-to-next-to-leading order corrections and higher-twist
contributions may have to be included to improve the theoretical
predictions.  Other possible explanations for the discrepancies
include non-perturbative string effects between the proton remnant and
the charm quark or an enhancement of the low-$x$ gluon component of
the photon, which is currently not well constrained experimentally.  A
similar picture has emerged in $\gamma\gamma$ collisions at
\mbox{LEP}. As shown by Andreev, the two-photon $D^*$ cross section is
slightly underestimated by NLO theory in the experimentally visible
region, in particular at low $p_t$ where resolved photon processes are
prominent.~\cite{andreev}

\begin{figure}[htb]
\vskip -0.25cm
\epsfig{figure=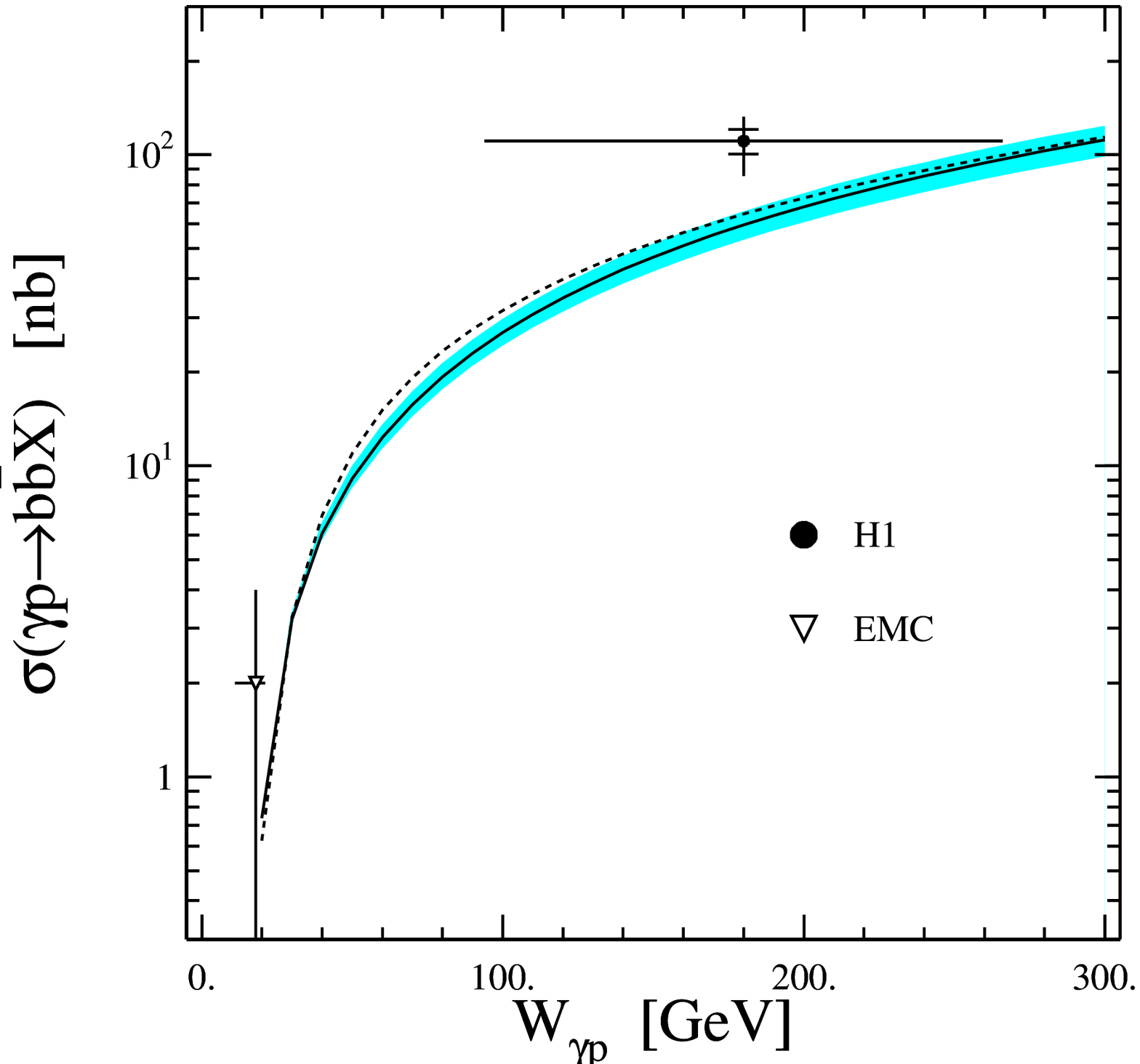,height=6cm}
\vskip -6.05cm
\hskip 5.9cm
\epsfig{figure=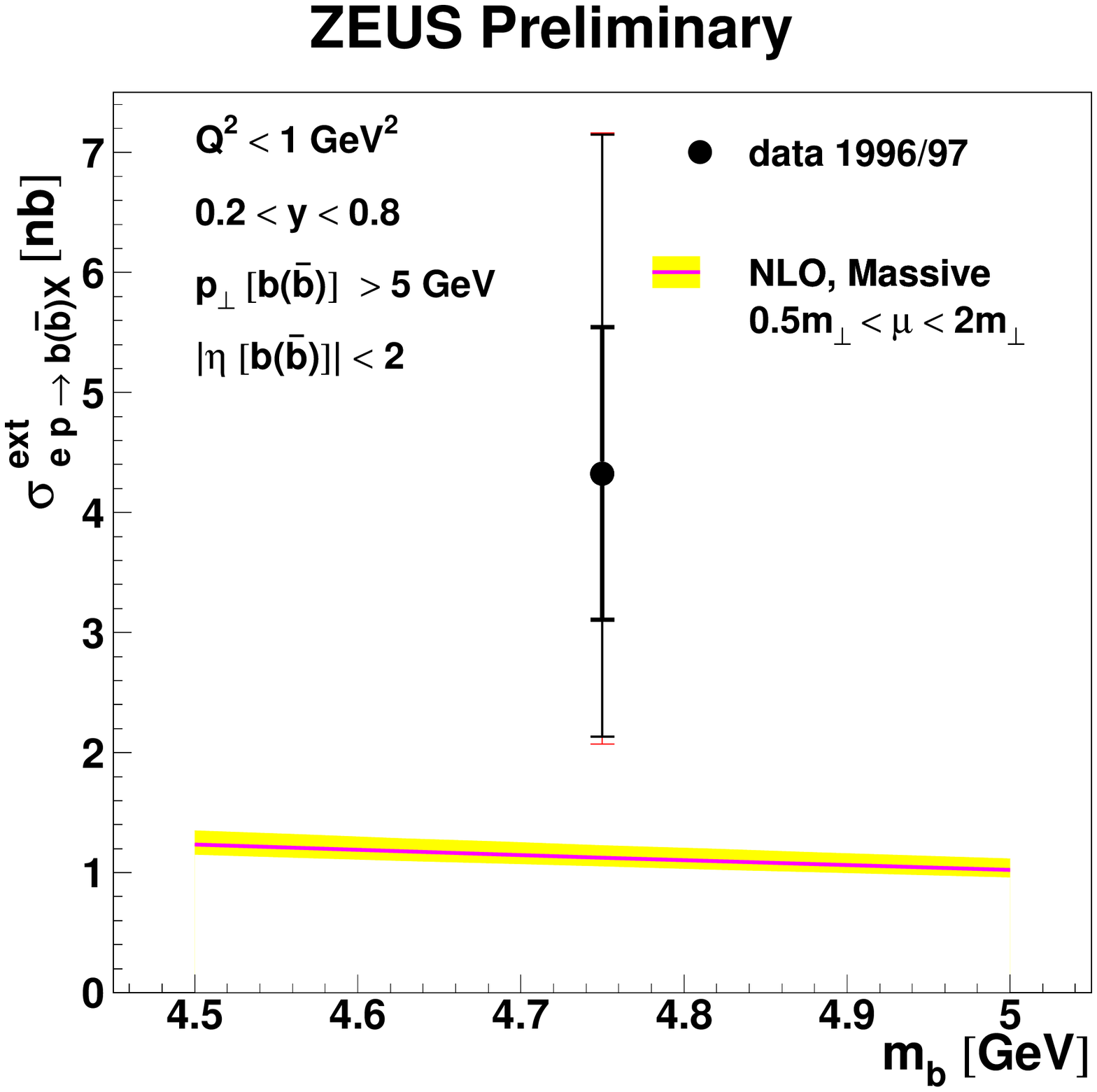,height=6cm}
\vskip -0.2cm
\caption[]{\label{fig:kuhr}
  Left: total photoproduction cross section for beauty production
  measured by H1 compared to the NLO QCD prediction. Right: cross
  section measured by ZEUS extrapolated to the parton level compared
  to the NLO QCD prediction.~\cite{kuhr}}
\vskip -0mm
\end{figure}

Due to the larger $b$ quark mass, theoretical predictions should be
under much better control for the beauty cross section.  Still, it is
well known that \mbox{Tevatron} data for the $b$ transverse momentum
distribution are systematically larger than NLO QCD predictions.  This
trend is supported by recent \mbox{HERA} and \mbox{LEP} measurements
which show that the beauty photoproduction cross section is
significantly higher than the NLO theory.~\cite{kuhr,andreev} Kuhr
presented the H1 and ZEUS results displayed in Fig.~\ref{fig:kuhr}.
Also the beauty cross section in $\gamma\gamma$ collisions at LEP as
measured by \mbox{L3}~\cite{andreev}
\begin{displaymath}
\sigma_{tot}(ee \to eeb\bar{b}X) = 
(9.9 \pm 2.9 \pm 3.8)~\mbox{pb} \qquad \mbox{[L3]}
\end{displaymath}
is higher than the NLO QCD calculation of $\approx 3$~pb.  The
discrepancy between NLO theory predictions and experiment in beauty
production should be taken seriously and possible theoretical
explanations have been discussed by Ellis in the final plenary
session.~\cite{ellis} The \mbox{HERA} luminosity upgrade and
alternative experimental techniques using microvertex-detection will
provide important additional information in the future.~\cite{kuhr}

\section{Charm production in deep-inelastic scattering}

The charm contribution to the total structure function $F_2$ at small
$x$ at \mbox{HERA} is sizeable, up to $25\%$.  A proper description of
the charm contribution to \mbox{DIS} is thus required for a global
analysis of structure function data and a precise extraction of the
parton densities in the proton.

Considerable theoretical effort has been devoted to including heavy
quark effects in \mbox{DIS}. As reviewed by Smith and Tung, different
so-called variable flavour number prescriptions have been defined in
the literature.~\cite{Smith:2000zc,tung} These prescriptions
incorporate the correct heavy quark threshold behavior and sum terms
$\propto \alpha_s \log{Q^2/M_c^2}$ when the hard scale $Q$ of the
reaction is significantly larger than the heavy-quark mass $M_c$.  The
structure functions calculated using different prescriptions generally
differ at finite order in perturbation theory, since the treatment of
the matching conditions and the extent to which finite mass effects
are retained in the coefficient functions is ambiguous.

Experimentally, charm production in \mbox{DIS} at \mbox{HERA} has not
only been studied for inclusive structure functions but differential
cross sections for $D^*$ production are available as
well.~\cite{gladilin} The predictions for differential distributions
in variable flavour number prescriptions at ${\cal O}(\alpha_s)$ are
spoiled by large scale dependences, with the exception of the $Q^2$
and $p_t(D^*)$ distributions where the agreement with data is
good.~\cite{tung} The scale dependence is significantly reduced when
higher-order corrections are included.  The comparison of \mbox{HERA}
data with ${\cal O}(\alpha_s^2)$ calculations~\cite{Smith:2000zc}
shows good overall agreement, with a possible slight excess of data in
the forward region.  Hadronization effects may well be responsible for
the deviation, similar to $D^*$ photoproduction.  A first measurement
of the charm contribution to the photon structure function has been
presented at this workshop and satisfactory agreement, within the
still sizeable statistical uncertainty, with theory was
found.~\cite{andreev}

\section{Quarkonium production}

Exciting phenomenological developments in quarkonium physics followed
from the application of non-relativistic QCD (NRQCD), an effective
field theory that includes the so-called colour-octet mechanisms.
Although gluon fragmentation into colour-octet charm quark pairs
appears as the most plausible explanation of the large direct $\psi$
production cross section observed at the \mbox{Tevatron}, the
applicability of NRQCD factorization to charmonium production is still
not established quantitatively.

Inclusive charmonium production at \mbox{HERA} offers unique
possibilities to assess the importance of different quarkonium
production mechanisms.  Kr\"uger showed that no conclusive evidence
for colour-octet processes has been observed in $J/\psi$
photoproduction or $J/\psi$ production in \mbox{DIS} so
far.~\cite{krueger} The measurements of $J/\psi$ production through
resolved photon processes, for which first results have been reported
at this workshop~\cite{krueger}, could be important to clarify the
issue once more statistics has been accumulated.

\begin{figure}[htb]
\begin{minipage}[htb]{.46\linewidth}
\epsfig{figure=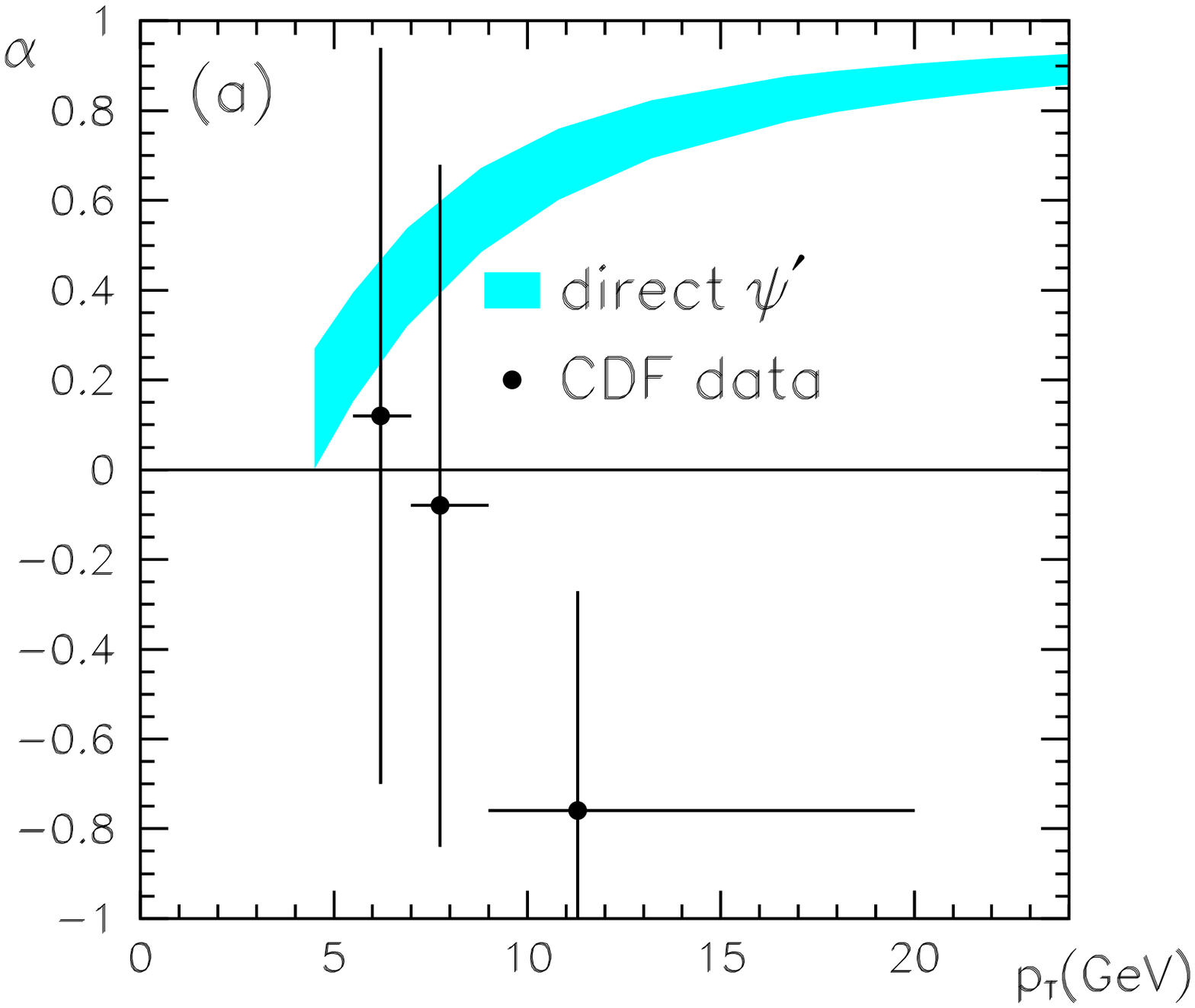,height=50mm,width=\linewidth}
\end{minipage}\hfill
\begin{minipage}[htb]{.46\linewidth}
\epsfig{figure=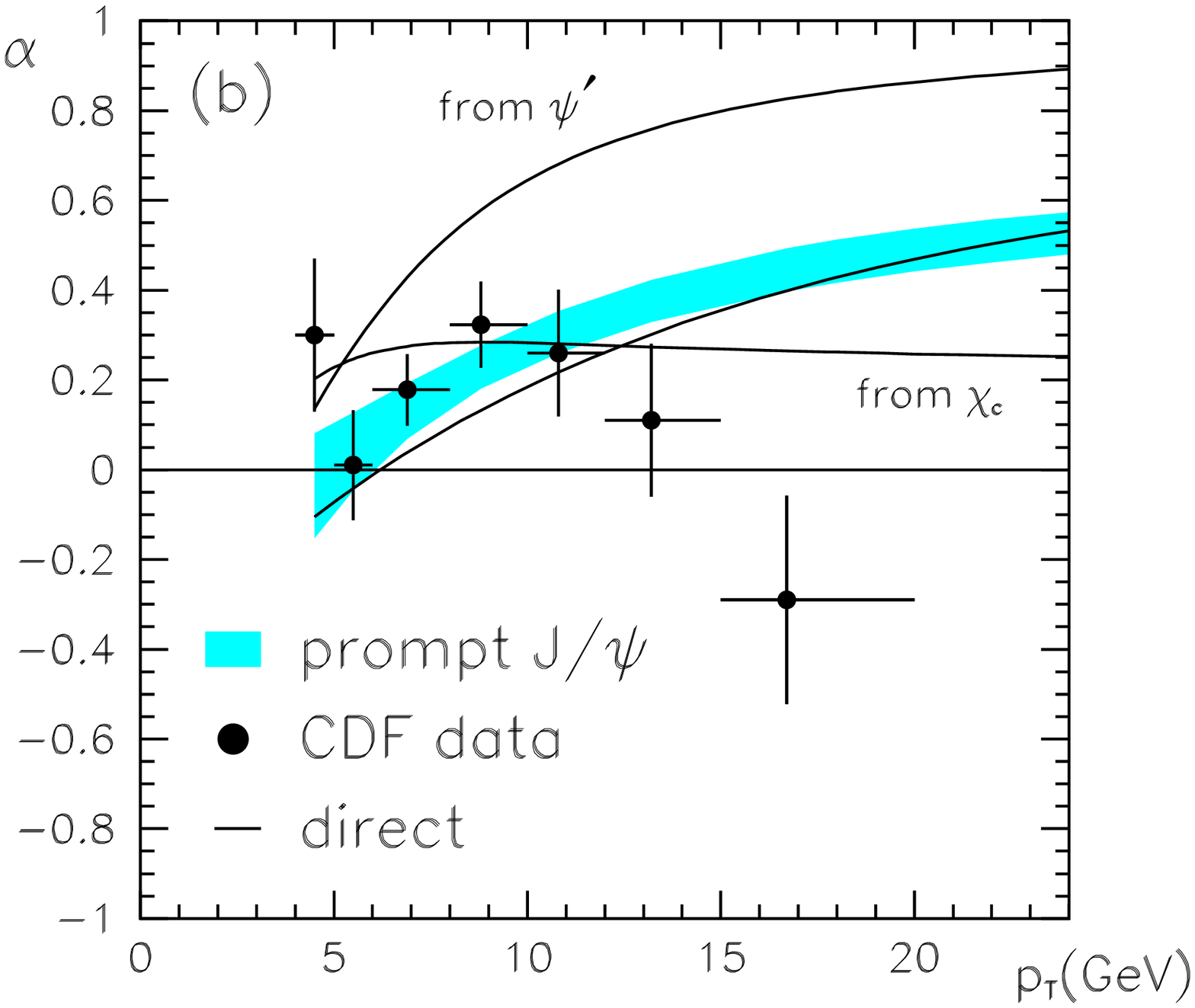,height=50mm,width=\linewidth}
\end{minipage}  
\caption[]{\label{fig:lee}
  Polarization variable $\alpha$ vs. $p_T$ for (a) direct $\psi'$ and
  (b) prompt $J/\psi$ compared to CDF data.~\cite{Lee:2000en} 100\%
  transverse polarization corresponds to $\alpha =1$.}
\vskip -0mm
\end{figure}

The single most crucial test of the NRQCD approach is the analysis of
direct $J/\psi$ and $\psi'$ polarization at large $p_t$ at the
\mbox{Tevatron}. NRQCD predicts a substantial fraction of transverse
polarization at $p_t$ above $\sim 10$~GeV, as discussed in Lee's
review.~\cite{Lee:2000en} As shown in Fig.~\ref{fig:lee} this
prediction is not supported by recent \mbox{Tevatron} data.  The
absence of transverse polarization in $J/\psi$ and $\psi'$
hadroproduction at large $p_t$, if confirmed by the higher statistics
data expected at Run~II at the \mbox{Tevatron}, would represent a
serious problem for the application of NRQCD to charmonium production.

\section{Discussion session} 

After 33 experimental and 12 theoretical presentations we ended our
working group meetings with a highly attended discussion session on
future directions in QCD at \mbox{HERA}.  Three `provocateurs',
M.~Wing, J.~Repond, and K.~Ellis started off the discussion.

Wing presented his views on the interesting directions in
photoproduction.~\cite{wing-web} Ellis expressed his discomfort with
the experimental results of too high cross sections for $b\bar
b$-production compared to current NLO QCD calculations, and he
discussed possible theoretical solutions.  Repond pointed out that
while most of us believe QCD to be {\em the theory} of strong
interactions, we do lack precision, i.e. $\alphas$ is only known to
about 3\%, we want to know where perturbative QCD is good enough,
where are higher order or resummation calculations needed and how to
do them correctly.  To pursue this experimentally, precision data and
wide coverage of the available phase space are needed.  The luminosity
upgrade of \mbox{HERA} will allow to reduce statistical and with much
effort by the experimentalists also systematic errors.  Also needed
are of course precision theory which already now is often trailing the
experiments.  Repond convincingly pleaded for an increase in the
manpower in theory to address, together with the experimentalists, the
outstanding QCD-issues at \mbox{HERA}.

We had an interesting discussion and good participation on the issues
mentioned above and some others brought up by the audience until we
ran out of time. Clearly we did not have enough time, probably we had
too many topics and not good enough structuring. The interest in such
discussions however could be seen from the lively debate we had and
the wishes for more by many people afterwards.

\section{Executive conclusions}

\begin{itemize}
\item
For many results, on jet cross sections, the determination of
$\alphas$, on charm cross sections, the major uncertainty is
theoretical due to the renormalization scale, followed by the
experimental error due to the hadronic energy scale of the
calorimeter.  Both need to be improved in the future.
\item
Using dijet rates instead of the dijet cross section, the error on
$\alphas$ due to the uncertainties in the parton density functions
could be halved.  In addition, this error can now be determined, since
parton densities with correlated errors have become available.
\item
Results on dijet production suggest that the parton densities of the
photon should be revised and improved by including \mbox{HERA} and
\mbox{LEP} data in a global fit.
\item
A first dedicated search for instantons in deep-inelastic scattering
was reported by H1.
\item
Several important pieces of next-to-next-to-leading order calculations
have been computed. The challenge is now to calculate the remainder
and to assemble the pieces.
\item
Resummed calculations for semi-inclusive hadron production and thrust
distributions at HERA are now available. Predictions for prompt photon
and jet production are still missing.
\item
Next-to-leading order QCD calculations for charm and beauty
photoproduction underestimate experimental data from \mbox{HERA} and
\mbox{LEP}. The discrepancy is particularly significant in the case of
beauty production where perturbative calculations should be on safe
grounds.
\item
In contrast, the comparison of \mbox{HERA} and \mbox{LEP} data on
charm production in DIS shows good overall agreement with ${\cal
O}(\alpha_s^2)$ calculations. A proper description of the charm
contribution to DIS is important for a precise extraction of the
parton densities in the proton.
\item
\mbox{HERA} measurements so far show no conclusive evidence for
colour-octet contributions to charmonium production in $\gamma p$ and
deep-inelastic reactions.  The absence of transverse polarization in
$J/\psi$ and $\psi'$ hadroproduction at large $p_t$ represents a
serious problem for the application of non-relativistic QCD to
charmonium production.
\item
Finally, one important echo from the discussion session of the last
day was that given the high experimental precision which has been
reached, more theoretical support is clearly needed.
\end{itemize}

\section*{Acknowledgments}
M.~Klasen is supported by Deutsche Forschungsgemeinschaft under
Contract KL 1266/1-1.  M.~Klasen and M.~Kr\"amer are supported in part
by the European Commission under Contract ERBFMRXCT980194.

\section*{References}

\end{document}